\documentclass[preprint,showpacs,superscriptaddress]{revtex4-1} 

\usepackage{amssymb}
\usepackage[dvips]{epsfig}
\usepackage{subfigure}
\usepackage{amsmath}

\usepackage{graphicx}
\usepackage{epstopdf}
\usepackage{color}
\usepackage{ulem}

\begin{document}

\title{The Asymmetric Active Coupler: \\
Stable Nonlinear Supermodes and Directed Transport}

\date{\today}

\author{Yannis Kominis}
\affiliation{School of Applied Mathematical and Physical Science, National Technical University of Athens, Athens, Greece}

\author{Tassos Bountis}
\affiliation{Departement of Mathematics, University of Patras,Patras, Greece}

\author{Sergej Flach}
\affiliation{Center for Theoretical Physics of Complex Systems, Institute for Basic Science, Daejeon, Korea}
\affiliation{New Zealand Institute for Advanced Study, Centre for Theoretical Chemistry \& Physics, Massey University, Auckland, New Zealand}

\begin{abstract}
We consider the asymmetric active coupler (AAC) consisting of two coupled dissimilar waveguides with gain and loss. We show that under generic conditions, not restricted by parity-time symmetry, there exist finite-power, constant-intensity nonlinear supermodes (NS), resulting from the balance between gain, loss, nonlinearity, coupling and dissimilarity. The system is shown to possess nonreciprocal dynamics enabling directed power transport and optical isolation functionality.    
\end{abstract}

\maketitle

Energy transport between coupled systems or different modes of the same system is one of the most fundamental problems in physics and the controlled and directed transport is of great importance in many technological applications such as electronic and optical devices. For the latter, the design and implementation of integrated photonic devices is a major challenge requiring the realization of a set of fundamental elements for photonic circuitry, such as couplers, switches, diodes and isolators for the directed transport of the optical power. \cite{Lifante}

The nonlinear coherent coupler \cite{Jensen_82, Daino_85} has been widely studied as a basic photonic component allowing for power-sensitive energy transport. It has been shown \cite{Chen_92,Thirstrup_95} that the presence of gain and/or loss in this system renders its dynamics more complex and enriches its functionality. Moreover, in the case where the gain in one channel is exactly equal to the loss in the other channel, the coupler can be considered as a $\mathcal{PT}$-symmetric dimer, and has been shown to possess unidirectional dynamics allowing for optical isolation functionality \cite{Ramezani_10, Sukhorukov_10} which is the key property for an optical diode. Similar properties have been studied for a large variety of such $\mathcal{PT}$-symmetric photonic structures, extending the theoretical interest on these systems, \cite{PT_theory} to realistic experimental studies on light propagation in coupled waveguide structures based either on AlGaAs heterostructures \cite{Guo_09} or on Fe-doped LiNbO$_3$ \cite{Ruter_10} at wavelenghts of 1550nm and 514.5nm, respectively. The $\mathcal{PT}$-symmetric systems have been considered for important applications such as the non-reciprocal light transmission \cite{Feng_11,Peng_14}, the observation of asymmetric transport, \cite{Bender_13} the study of active coupling mechanisms, \cite{Barashenkov_14} and the synthesizing of unidirectionally invisible media \cite{Lin_11}.  Also, $\mathcal{PT}$-symmetric cavities have been studied with respect to interesting properties of resonant mode control and selection, which is of crucial importance in laser physics. \cite{lasers}  
The presence of gain and loss along with the nonlinearity of a photonic structure has also been shown to support bright and dark solitons in dual-core systems \cite{Malomed} and to provide soliton control capabilites in photonic structures with homogeneous gain and loss \cite{Kominis_homogeneous} as well as in structures with symmetric \cite{Musslimani_08} or nonsymmetric \cite{Kominis_inhomogeneous} spatially inhomogeneous gain and loss.  
Finally, we stress the relation of the underlying model of active photonic structures, consisting of coupled mode equations, with similar models used in the study of quantum systems including Bose-Einstein and exciton-polariton condensates \cite{Aleiner_12,Rayanov_15,Rahmani_16}.

The $\mathcal{PT}$-symmetric dimer is known to generate unstable dynamics above the parameter threshold which separates the $\mathcal{PT}$ invariant dynamics from the $\mathcal{PT}$-broken dynamics
\cite{Barashenkov_13}.
One way to regain stability is to use the analogy to dissipatively coupled exciton-polariton condensates in the weak lasing regime \cite{Aleiner_12,Rayanov_15}. In the optical coupler case this implies to
place an active
medium in the evanescent wave region of the coupler \cite{Barashenkov_14} This is a rather complicated and intricate experimental task, because the pumping in the evanescent wave region can easily
lead to an overpumping, which will substantially modify the used underlying model equations. 

In this work, we investigate a much more straightforward and simpler way by raising the restriction of any symmetry in a system with gain and loss. More specifically, we study the most general case of a nonlinear Asymmetric Active Coupler (AAC) where the two constituents can be dissimilar and can also have arbitrary gain and loss. The dynamics of the system is shown to possess stable transport features and capabilities for optical isolation functionality. The dynamical regimes are crucially depending on the existence of constant-intensity nonlinear supermodes (NS) resulting from the dynamical balance between the effects of nonlinearity, coupling, gain and loss. The existence of stable NS allows for the directed power transfer. Surprisingly, it is the absence of symmetry, that enables the existence of finite-power modes of the system, in contrast to the $\mathcal{PT}$-symmetric case \cite{Ramezani_10, Barashenkov_13, Kevrekidis, Konotop} where no such modes exist and the undesirable effect of unbounded power increasing takes place. The freedom in the selection of the system parameters, provides potential for multifunctional capabilities of the AAC as a basic component for integrated photonic circuitry.

For an Asymmetric Active Coupler (AAC), the modal amplitudes of the two individual waveguides are governed by the coupled mode equations
\begin{eqnarray}
 -i\frac{dA_1}{dz}&=&(\beta_1+i\alpha_1)A_1+\gamma\left(|A_1|^2+\sigma|A_2|^2\right)A_1+\frac{\kappa}{2}A_2 \label{CM_1}\;,\\
 -i\frac{dA_2}{dz}&=&(\beta_2+i\alpha_2)A_2+\gamma\left(\sigma|A_1|^2+|A_2|^2\right)A_2+\frac{\kappa}{2}A_1 \label{CM_2}\;,
\end{eqnarray}
where $\beta_j+i\alpha_j$ is the complex propagation constant of waveguide $j$ with $a_j>0 (<0)$ referring to loss (gain), $\kappa/2$ is the linear coupling coefficient and $\gamma$, $\sigma$ are the nonlinear SPM and XPM parameters, respectively \cite{Chen_92}.
Let us introduce the Stokes parameters: $S_0=|A_1|^2+|A_2|^2, S_1=|A_1|^2-|A_2|^2, S_2=A_1^*A_2+A_1A_2^*, S_3=i(A_1^*A_2-A_1A_2^*)$.
While $S_0$ measures the total power in the coupler, the component $S_1$ quantifies the deviation from an exact power balance in both waveguides.
The coupled mode equations (\ref{CM_1})-(\ref{CM_2}) can then be written as
\begin{eqnarray}
 \frac{dS_0}{dz}&=&-\delta S_0-\alpha S_1 \label{dS0}\;, \\
 \frac{dS_1}{dz}&=&-\alpha S_0-\delta S_1+\kappa S_3 \label{dS1}\;,\\
 \frac{dS_2}{dz}&=&-\delta S_2-(\beta+\chi S_1) S_3 \label{dS2}\;,\\
 \frac{dS_3}{dz}&=&-\delta S_3+(\beta+\chi S_1)S_2-\kappa S_1 \;. \label{dS3}
\end{eqnarray}

We consider cases where $\alpha_1 \alpha_2 <0$ so that the sign of the parameter $\alpha=\alpha_1-\alpha_2$ determines whether the first waveguide has loss and the second has gain ($\alpha>0$) or vice versa ($\alpha<0$).
The crucial parameters $\delta=\alpha_1+\alpha_2$ and $\beta=\beta_1-\beta_2$ determine the excess gain/loss and the asymmetry of the coupler and are quantifying the deviation from the $\mathcal{PT}$ symmetry point at which $\delta=\beta=0$. Finally $\chi=\gamma(1-\sigma)$. From the definition of the Stokes parameters it follows that $S_0^2=S_1^2+S_2^2+S_3^2$
so that the dynamics of the system of differential equations (\ref{dS0})-(\ref{dS3}) essentially takes place in a three-dimensional and is described by the Stokes vector $\vec{S}\equiv(S_1,S_2,S_3)$, governed by Eqs. (\ref{dS1})-(\ref{dS3}). 

In the absence of gain and loss $(\delta=\alpha=0)$, the total power is conserved so that $S_0$ is invariant. Moreover, in this case there exists an additional invariant $\Gamma=S_2+(\beta/\kappa)S_1+(\chi/2\kappa)S_1^2$. The dynamics of the system is integrable and can be completely described in terms of these two invariants \cite{asymmetric_dimer}. 
For the symmetric coupler $(\beta=0)$ it has been shown that system dynamics is similar to that of a Duffing equation,  that has stable and unstable fixed points which correspond to Nonlinear Supermodes,and nonharmonic periodic orbits for the evolution of the Stokes vector \cite{Jensen_82, Daino_85}. The dynamics of the system is reciprocal with respect to initial conditions corresponding to symmetric power distribution in the two waveguides.     

The presence of gain and loss renders the dynamics of the system non-integrable in general and results in complex dynamics \cite{Chen_92, Thirstrup_95}. 
In the special case where the coupler is $\mathcal{PT}$-symmetric $(\delta=0)$, there exist two new invariants of motion, rendering the system integrable, despite the fact that the total power $S_0$ is not conserved \cite{Ramezani_10, PT_integrable, Kevrekidis}. 
The dynamics of the system is non-reciprocal. However, no finite-power Nonlinear Supermodes exist since the total power of the system continuously increases or decreases, depending on the sign of $\alpha$.

The introduction of asymmetry in the structure allows for the existence of fixed points of the system of differential Eqs. (\ref{dS1})-(\ref{dS3}) which correspond to  finite-power, constant-intensity Nonlinear Supermodes of the AAC. These supermodes represent optical fields that propagate unchanged along the coupler  despite of the presence of gain, loss, asymmetry and nonlinear effects. 
They are obtained by zeroing the left hand sides of equations (\ref{dS1})-(\ref{dS3}). We first note that there exists always a trivial zero fixed point $O$ for which $S_1=S_2=S_3=0$.


In order to find the nontrivial supermodes, it is useful to define the normalized Stokes vector as $\vec{F}=\vec{S}/S_0$. Then the nonzero fixed points for any set of parameter values are located on the surface of a Bloch sphere of unit radius and are given by $\vec{F}_{\pm}^{(0)}=(F_1^{(0)},F_2^{(0)},F_3^{(0)})$ with 
\begin{eqnarray}
 F_1^{(0)}&=&-\Delta\;,\\
 F_2^{(0)}&=&\mp \frac{K}{2}\sin\phi\;,\\
 F_3^{(0)}&=& \frac{K}{2}\left(1-\cos\phi\right) \;,
\end{eqnarray}
where $\tan\left(\phi/2\right) \equiv |\Delta / \Lambda_\pm|$, $\Lambda_\pm  = \pm \Delta \sqrt{(K^2+\Delta^2-1)/(1-\Delta^2)}$.
Surprisingly the location of the nonlinear supermodes are described by only two parameters $\Delta$ and $K$ on the unit Bloch sphere. These parameters are obtained in units of the parameter $\alpha$ as $\Delta \equiv \delta/\alpha $,  $K \equiv \kappa / \alpha $. In addition, the normalized nonlinearity parameter $X\equiv \chi / \alpha$ enters the conditions for the existence of the nonlinear supermodes: 
\begin{eqnarray}
 1-K^2 \leq \Delta^2 &\leq& 1 \label{C1}\\
 \Delta X (B-\Lambda_\pm)&\geq& 0 \;. \label{C2}
\end{eqnarray}

As shown in Fig. 1, the location of the Nonlinear Supermodes for different $\Delta$ depends on whether $K$ is greater or less than unity, and all curves touch at their common points at $\vec{F}=(\pm 1,0,0)$. The two Nonlinear Supermodes (corresponding to different signs of $F_2^{(0)}$) are symmetric with respect to the plane $F_2=0$.  Opposite values of $\Delta$ and $K$ result in fixed points symmetric with respect to the planes $F_1=0$ and $F_3=0$. The value of $F_1$ is of particular importance since it is directly related to the ratio of modal amplitudes of the two waveguides. For $F_1>0$ ($F_1<0$) the modal amplitude of the first (second) waveguide is larger and as $F_1 \rightarrow 1 (-1)$ all the power tends to concentrate on the first (second) waveguide. The total power cannot be located in a single waveguide as long as $\alpha_1,\alpha_2 \neq 0$. Therefore we have $|F_1^{(0)}|=|\Delta|<1$ and there is always nonzero power in both waveguides. However, appropriate parameter selection can reduce the power in one of the waveguides at any desirable level, resulting in sufficient optical isolation. Moreover, it is readily shown from the sign of $F_1^{(0)}$ that, for the case of net loss ($\alpha_1+\alpha_2>0$) most of the power is located at the waveguide with gain, whereas for the case of net gain ($\alpha_1+\alpha_2<0$) most of the power is located in the lossy waveguide.

The existence of the Nonlinear Supermodes depends on the parameters of the structure through the conditions (\ref{C1}) and (\ref{C2}). For $|K|>1$, the NS exist for $\Delta$ lying in the value range $|\Delta|<1$ whereas for $|K|<1$ the NS exist for the two disjoint value ranges defined by $\sqrt{1-K^2}<|\Delta|<1$. 
In terms of the waveguide parameters, condition (\ref{C1}) is writen as $-\kappa^2/4 \leq \alpha_1 \alpha_2 \leq 0$, implying that, for the existence of a Nonlinear Supermode, it is necessary to have one waveguide with gain and one with loss (not necessarily of equal amplitude), as expected, in order to have some power balance. Note that, this condition is much less restrictive than symmetry conditions such as in the $\mathcal{PT}$-symmetric couplers.
The existence of the two NS also depends crucially on the value of $B=(\beta_1-\beta_2)/(\alpha_1-\alpha_2)$ as shown from the condition (\ref{C2}), so that, depending on the degree of asymmetry of the two waveguides, there exist either two, one or zero NS. More specifficaly, for $\Delta X >0$ the Nonlinear Supermodes $\vec{F}_+^{(0)}$ and $\vec{F}_-^{(0)}$ exists for $B > \Lambda_+$ and $B > \Lambda_-$, respectively, whereas for $\Delta X < 0$ the inequalities for $B$ are reversed. The total power of the two nonlinear supermodes is given by $S_{0,\pm}=(B-\Lambda_\pm)/\Delta X$ so that $S_{0,-}>S_{0,+}$ for $\Delta X > 0$, whereas the opposite holds for $ \Delta X < 0$. 

The stability of the Nonlinear Supermodes is determined by the eigenvalues of the Jacobian matrix of the system (\ref{dS1})-(\ref{dS3}). The domains of existence of the Nonlinear Supermodes $(\vec{F}_{\pm}^{(0)})$ as obtained by the conditions (\ref{C1}) and (\ref{C2}) in the $(\Delta,B)$ parameter space are shown along with their stability type in Fig. 2 for  $\alpha=1$ and $X=1$, $K>1$ (a) and $K<1$ (b). For $K>1$ there exist a stable NS for every value of $|\Delta|<1$ [Fig. 2(a)] in contrast to the case where $K<1$ [Fig. 2(b)]. In both cases a stable NS exists in parameter regions where $\Delta B>0$. Both NS bifurcate from the zero state $S_0=0$ at the points $B=\Lambda_{\pm}$ with eigenvalues $\lambda=0,\alpha(-\Delta+i\Lambda_{\pm}/\Delta)$. 


It is worth noticing that the system of coupled mode equations (\ref{CM_1}),(\ref{CM_2}) is invariant under the ``staggering'' transformation $\gamma \rightarrow -\gamma$, $A_1 \rightarrow -A_1^*$, $A_2 \rightarrow A_2^*$, $\beta_{1,2} \rightarrow -\beta_{1,2}$. Therefore, the existence and stability of the NS for a defocusing nonlinearity ($\gamma, X < 0$) can be directly determined from the case of a focusing nonlinearity ($\gamma, X > 0$) by inverting the signs of $\beta_{1,2}$, and, in that sense, the two cases are dynamically equivalent. This is in contrast to the case of an actively coupled dimer where the two cases undergo different dynamics, and the defocusing case has blow-up regimes. \cite{Barashenkov_13}   

Apart from the nonzero fixed points corresponding to NS, for any parameter set (including symmetric and nonsymmetric cases) there exists a zero fixed point $O$ of Eqs.(\ref{CM_1})-(\ref{CM_2}) corresponding to a trivial (zero) state with eigenvalues and eigenvectors given by $\lambda_{1,2}=\frac{\alpha}{2}\left(\Omega+i\Delta\pm\sqrt{(B+i)^2+K^2}\right)$ and
$\vec{e}_{1,2}=\left[\left(-(B+i)/K\pm \sqrt{[(B+i)/K]^2+1}\right)^{-1},1\right]$, where $\Omega=(\beta_1+\beta_2)/\alpha$. The dependence of the stability of the trivial (zero) state $O$ on the parameters of the AAC is also depicted in Fig. 2.

The asymmetry of the structure allows for nonreciprocal dynamics and directed transfer of power between the two waveguides. In the following we investigate the dynamics of the system for the initial conditions $\vec{S}\equiv(S_1,S_2,S_3)=(\pm1,0,0)$ corresponding to the cases where power is initially launched exclusively in one of the two waveguides. The case of an AAC with parameters corresponding to Fig. 2(b) for various values of $B$ is investigated in Fig. 3. For $B=0.8$ it is shown [Fig. 3(a)] that both initial conditions result in an asymmetric distribution of power between the two waveguides, corresponding to the stable NS with $S_1/S_0=-\Delta=-0.7$. No matter in which waveguide the initial power is injected, the system evolves to a stable state where the ratio of the modal amplitudes in the two waveguides is $|A_1|^2/|A_2|^2=(1-\Delta)/(1+\Delta)=0.18$ and the total power is $S_{0,+}=0.64$. For $B=0.2$, as shown in Fig. 2(b), only the unstable NS exists. The initial condition $\vec{F}=(1,0,0)$ evolves to the zero state (center of the Bloch sphere) whereas the initial condition $\vec{F}=(-1,0,0)$ evolves to a state of continuously increasing $S_0$ (blow up solution) corresponding to the point $(-1,0,0)$ of the Bloch sphere, as shown in Fig. 3(b). Finally, for $B=-0.2$ no Nonlinear Supermode exists and both initial conditions evolve to the state of continuously increasing $S_0$  (blow up solution) corresponding to the point $(-1,0,0)$ of the Bloch sphere, as shown in Fig. 3(c), similarly to the case of a $\mathcal{PT}$-symmetric coupler. \cite{Ramezani_10} In all three cases the dynamics of the system is nonreciprocal and directed power trasnfer takes place. However, only under parameter values for which a stable Nonlinear Supermode exists, the system evolves to a final state of finite total power.

The above feature can be further exploited for the operation of the AAC as an optical isolator. The optical isolation for a state corresponding to a Nonlinear Supermode is directly determined by the parameter $\Delta$. From the conditions for the existence of a stable NS (\ref{C1}) and (\ref{C2}), as depicted in Fig. 2, it is shown that as $|\Delta|$ approaches unity - corresponding to perfect isolation - an increasing value of $B$ is required. For a value $\Delta=0.95$ both initial conditions $\vec{F}=(\pm1, 0, 0)$ evolve to a final state with a ratio of modal amplitudes $|A_1|^2/|A_2|^2=0.026$, as shown in Fig. 4, so that independently of the waveguide in which the power is initially launched the system evolves to a final state where the total power is finite and almost all power is located in the second waveguide. Note that the amount of power remaining in the first waveguide can be set as small as desired, by choosing a $\Delta$ close to unity and appropriate values for $B$ and $K$. It is worth emphasizing that the final state has a finite total power $S_0=1.85$ for $B=4$, in contrast to $\mathcal{PT}$-symmetric couplers where the final system state corresponding to optical isolation has continuously increasing total power. \cite{Ramezani_10} As shown in Fig.4(a), the initial condition corresponding to power injected exclusively at the first waveguide leads to an evolution according to which the system initially approaches the unstable trivial solution (zero fixed point) and subsequently evolves to the stable nonzero fixed point corresponding to the Nonlinear Supermode.

In conclusion, we have investigated new posibilities for directed transport in active structures, opened by raising the restriction of spatial symmetry of the conservative and non-conservative properties of the system. For the case of an Asymmetric Active Coupler, it has been shown that it is the absence of symmetry that allows for the existence of finite-power Nonlinear Supermodes that can be utilized for power transport control and optical isolation. The results are directly applicable to any type of active dimer such as coupled cavities and electronic circuits in the realm of physics as well as in chemistry and biology applications where similar coupled mode models are used, and can also be generalized for oligomers and networks.

\clearpage
\begin{figure}[pt]
  \begin{center}
  {\scalebox{0.4}{\includegraphics{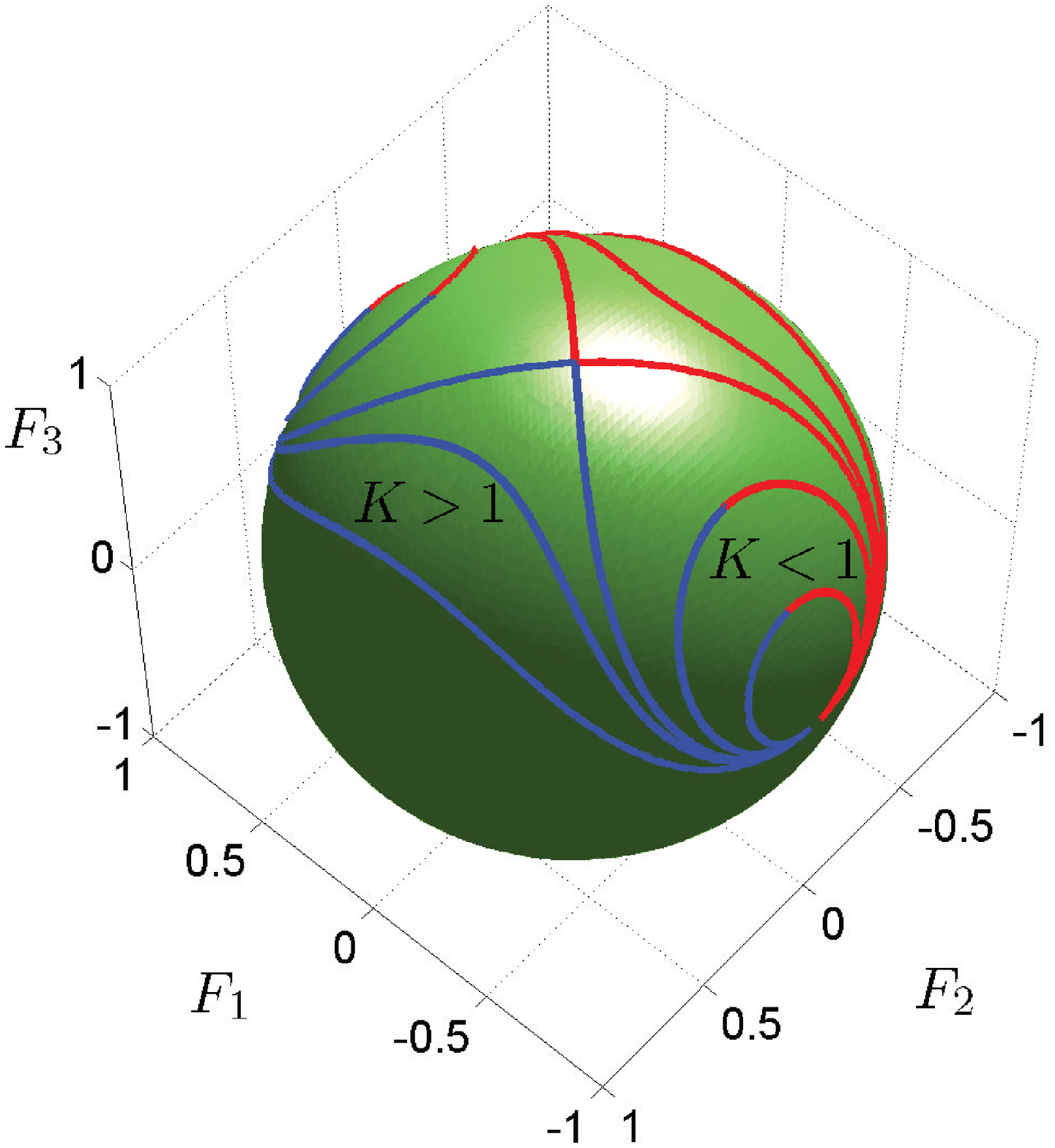}}}
  \caption{The location of the two Nonlinear Supermodes $(\vec{F}_{\pm}^{(0)})$ (red/blue curves) of the Asymmetric Active Coupler on the surface of a Bloch sphere of unit radius for different values of $K>0$ and $\Delta$. Different curves correspond to given values of $K$ and varying $\Delta$. The topology of the curves depends drastically on whether $K$ is greater or less than unity. For $K=1$ the curves intersect at $\vec{F}=(0,0,1)$. All curves are tangent at $\vec{F}=(\pm1,0,0)$.}
  \end{center}
\end{figure}

\begin{figure}[pt]
  \begin{center}
  \subfigure[]{\scalebox{0.2}{\includegraphics{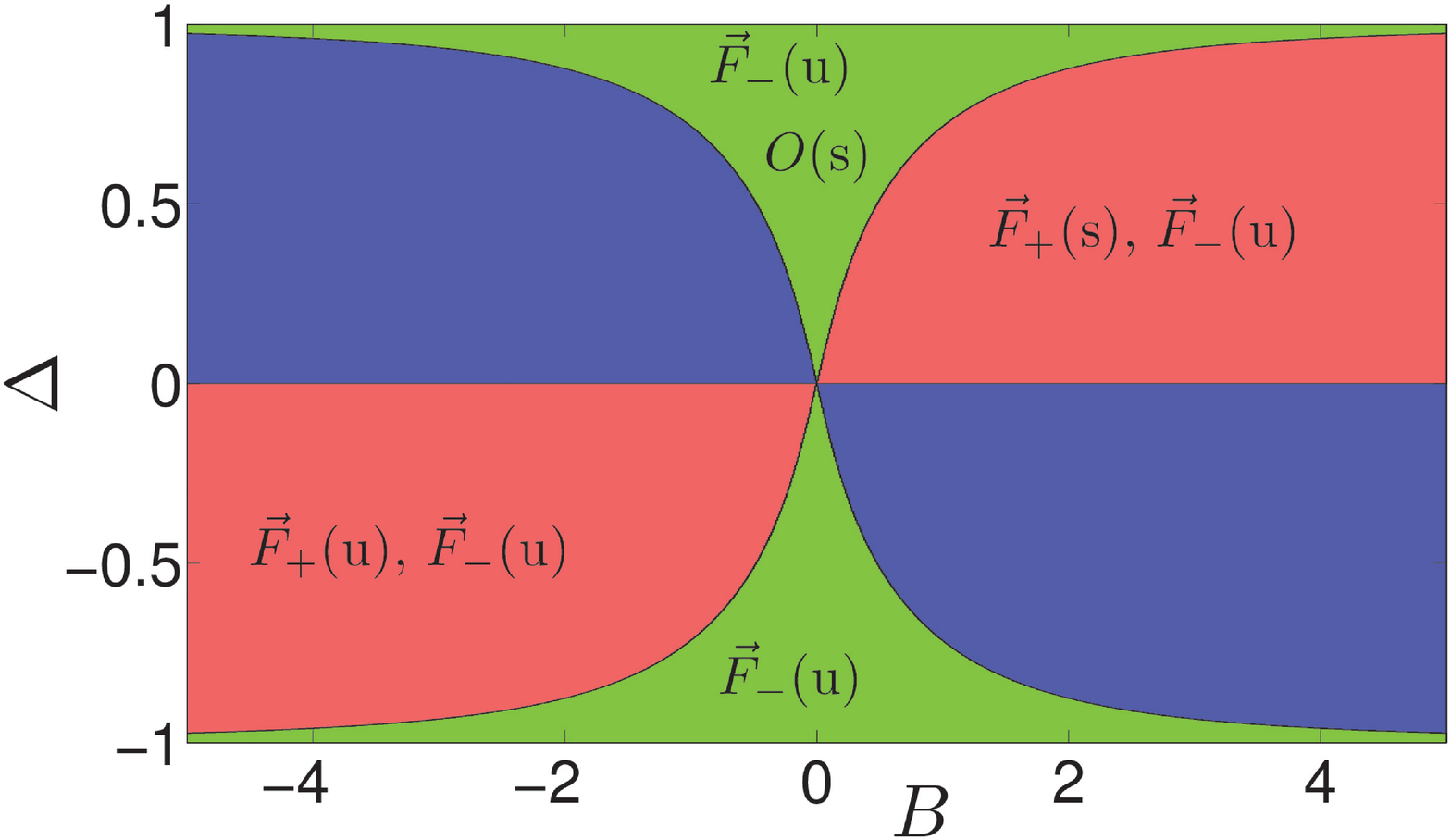}}}
  \subfigure[]{\scalebox{0.2}{\includegraphics{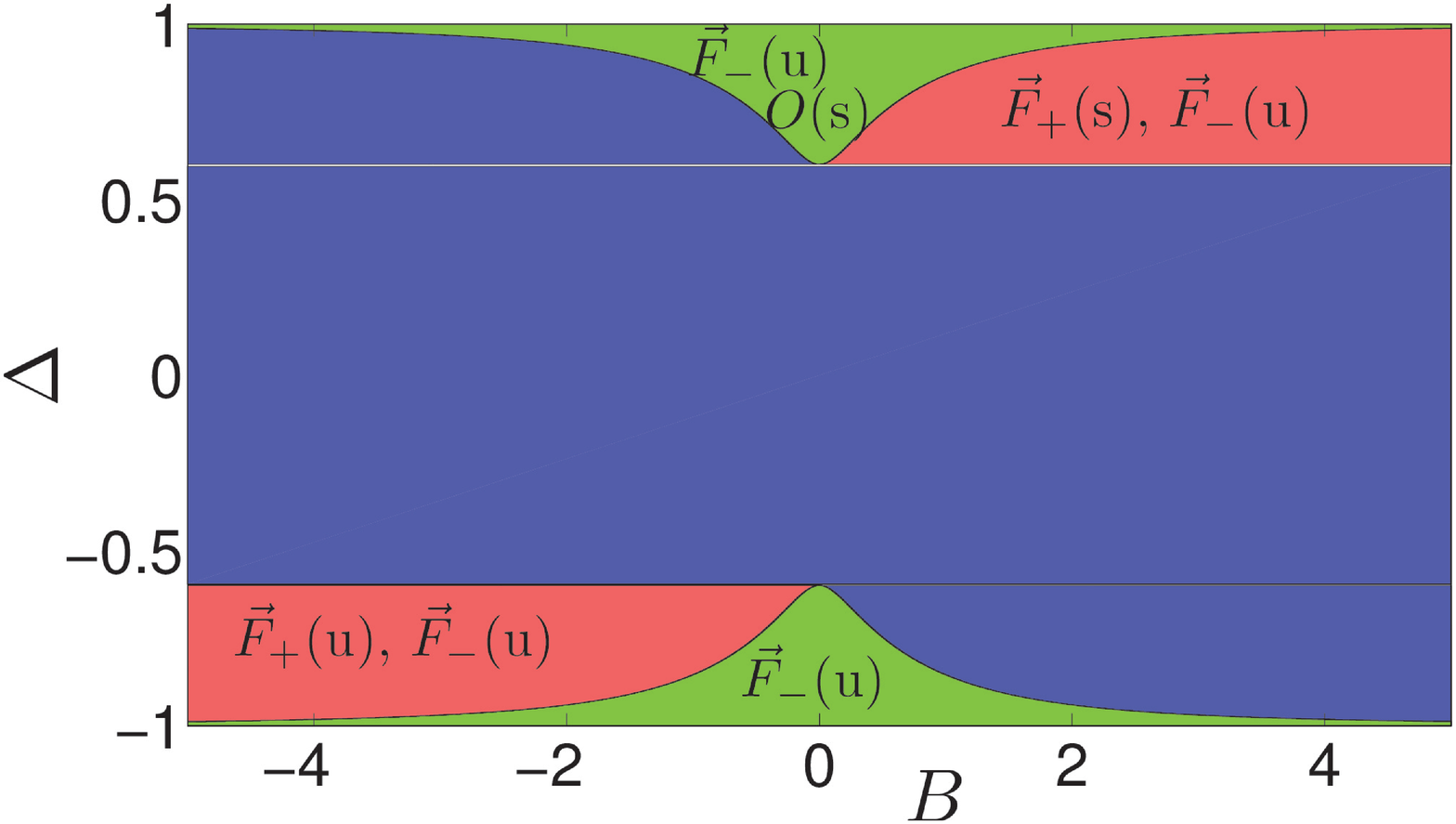}}}
  \caption{The domains of existence of stable (s) and unstable (u) Nonlinear Supermodes $(\vec{F}_{\pm}^{(0)})$ of the Asymmetric Active Coupler in the $(\Delta,B)$ parameter space for $\alpha=1$ and $X=1$. (a) $K=1.2$, (b) $K=0.8$. The zero fixed point $O$ exists for all parameter values but it is stable only in the regions marked with $O(s)$.}
  \end{center}
\end{figure}


\begin{figure}[pt]
  \begin{center}
  \subfigure[]{{\scalebox{0.2}{\includegraphics{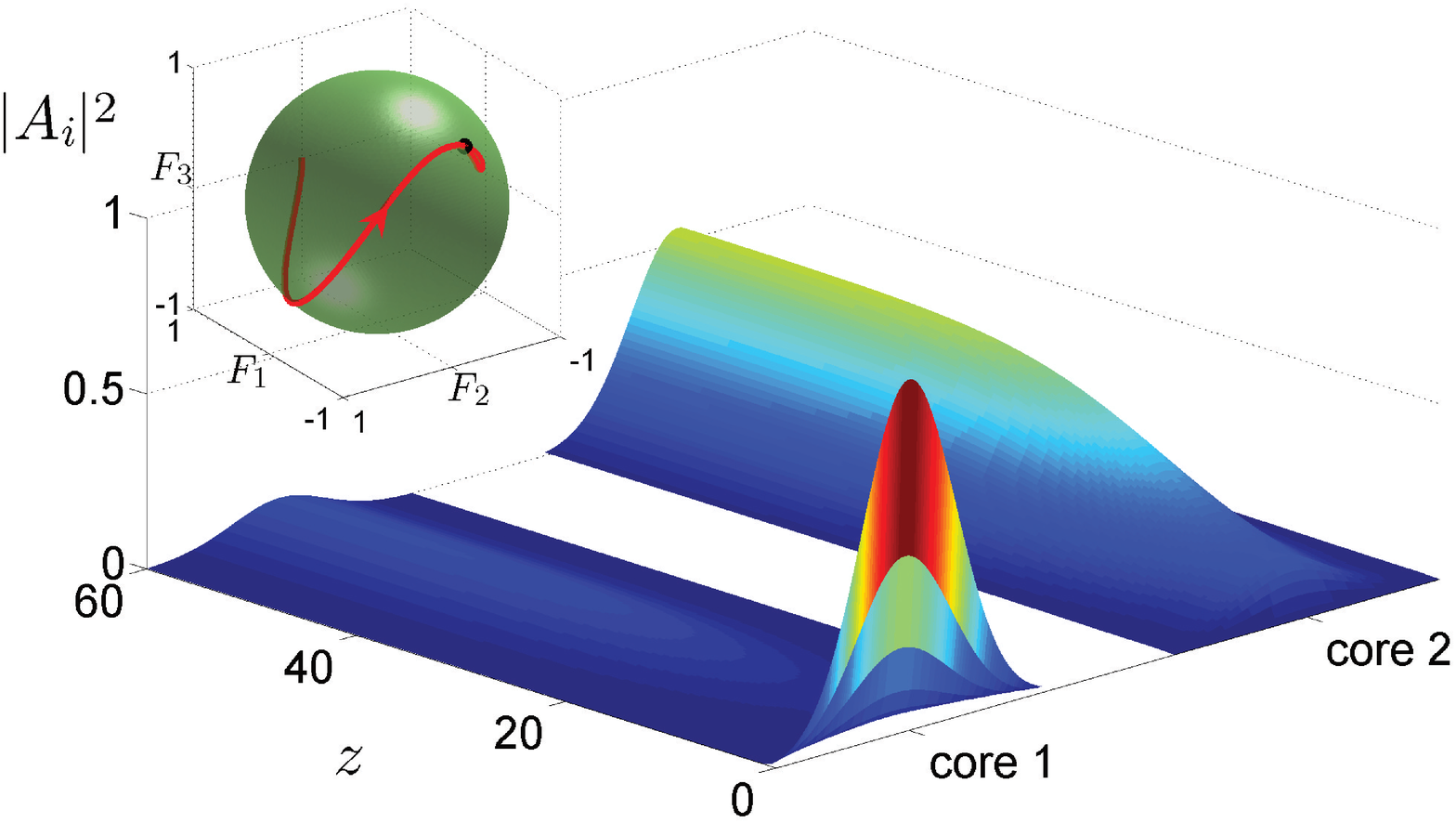}}}{\scalebox{0.2}{\includegraphics{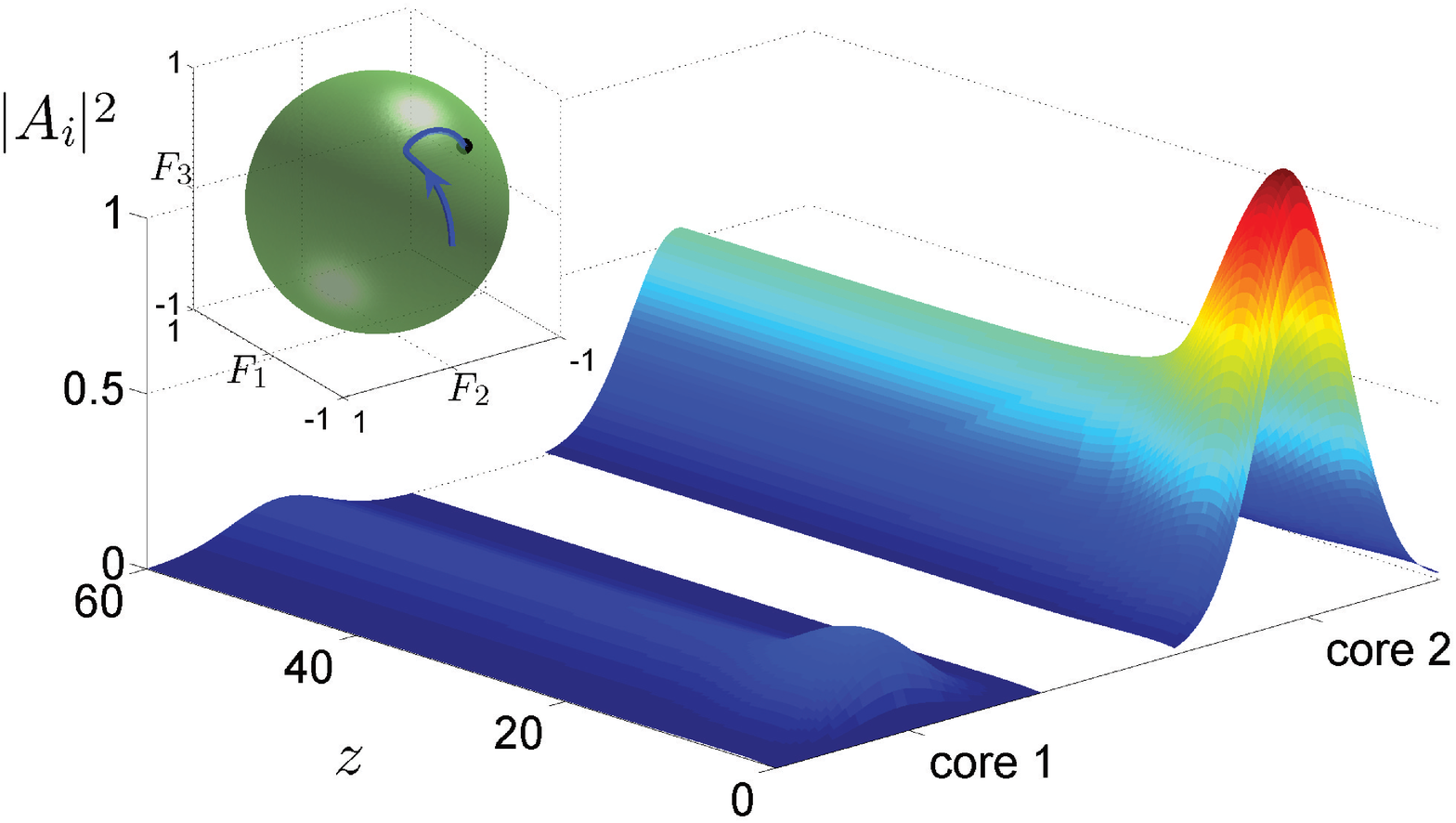}}}}\\
  \subfigure[]{{\scalebox{0.2}{\includegraphics{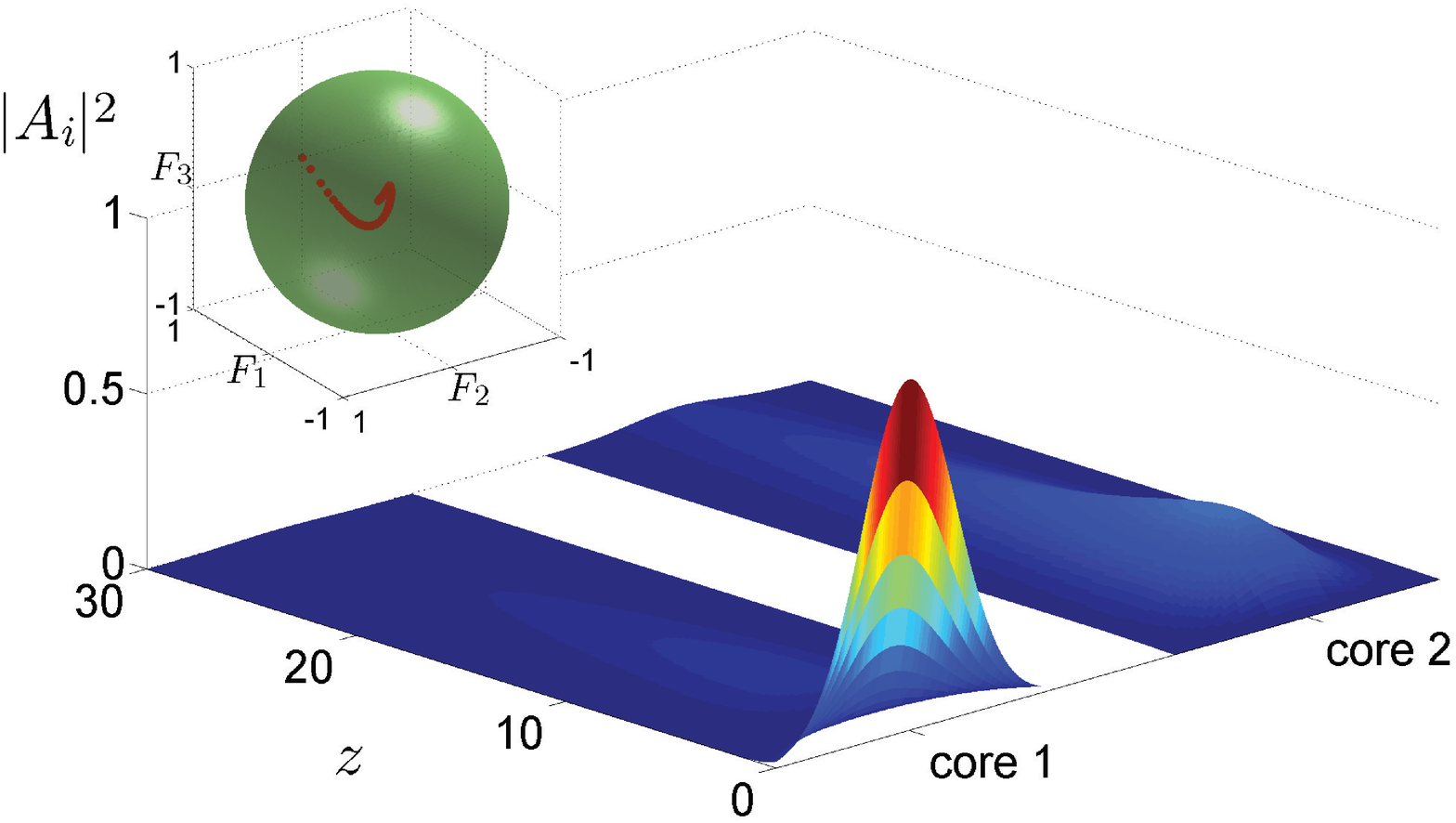}}}{\scalebox{0.2}{\includegraphics{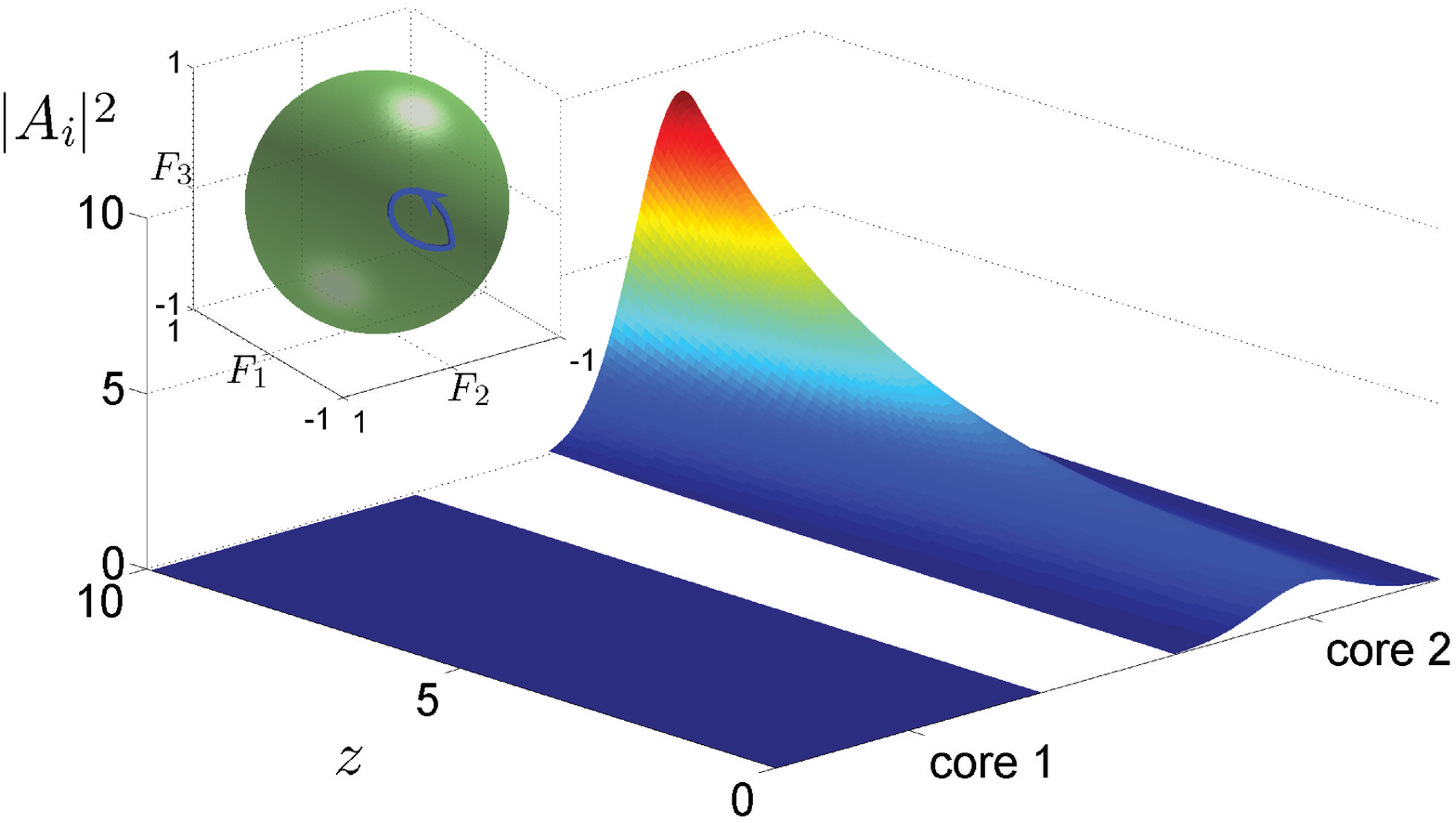}}}}\\
  \subfigure[]{{\scalebox{0.2}{\includegraphics{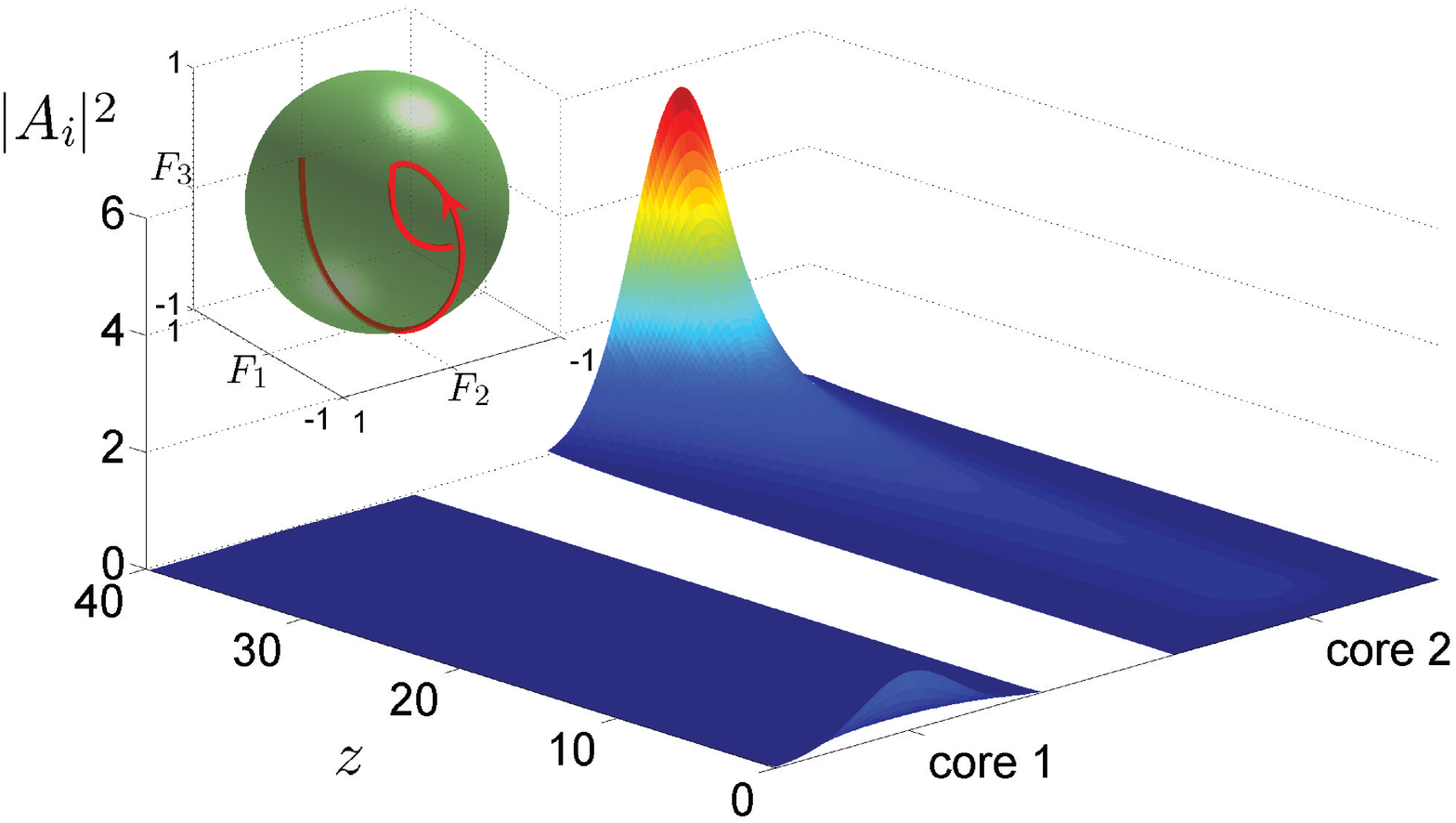}}}{\scalebox{0.2}{\includegraphics{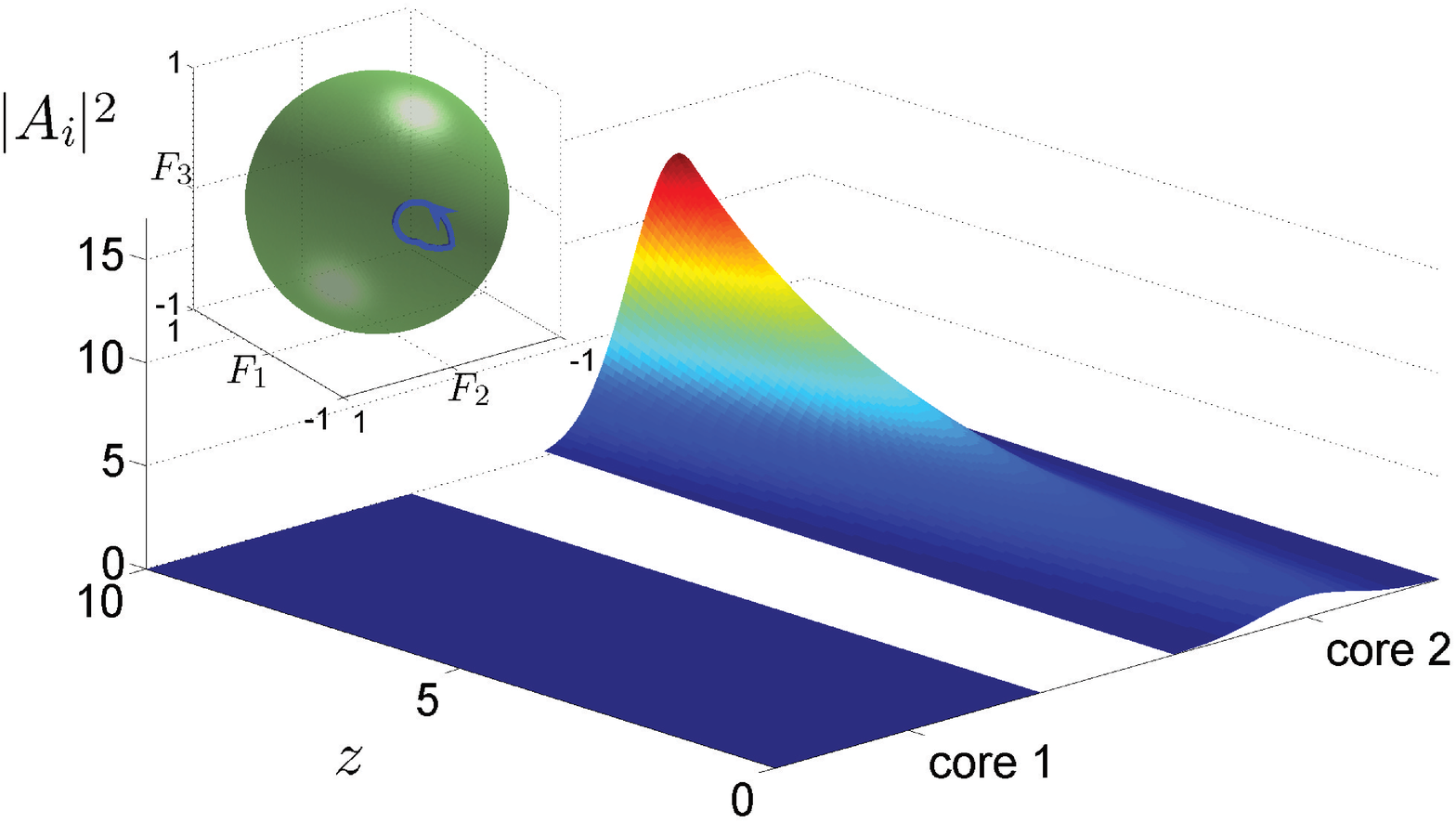}}}}
  \caption{Nonreciprocal dynamics of an Asymmetric Active Coupler with $\alpha=1$, $X=1$, $K=0.8$, $\Delta=0.7$ and $B=0.8$(a), $B=0.2$(b), $B=-0.2$(c). Initial conditions corresponding to initial power injected exclusively in one of the two waveguides are located at the poles of the Bloch sphere $\vec{F}=(\pm1,0,0)$ (red/blue curves). The insets depict the evolution of the Stokes vector on a Bloch sphere of unit radius. (a) Existence of a stable NS; the trivial (zero) fixed point is unstable: No matter in which waveguide the power is initially launched, the final power distribution in the two waveguides is determined by the stable NS. (b) Existence of an unstable NS; the trivial (zero) fixed point is stable: When the power is launched in the first waveguide it evolves to the trivial state (red dotted curve), whereas when power is launched in the second waveguide, it evolves to an unbounded (blow up) state where power is located in the second waveguide. (c) Nonexistence of a NS; the trivial (zero) fixed point is stable: No matter in which waveguide the power is initially launched it evolves to the unbounded (blow up) state where power is located in the second waveguide.}
  \end{center}
\end{figure}

\begin{figure}[pt]
  \begin{center}
  \subfigure[]{\scalebox{0.2}{\includegraphics{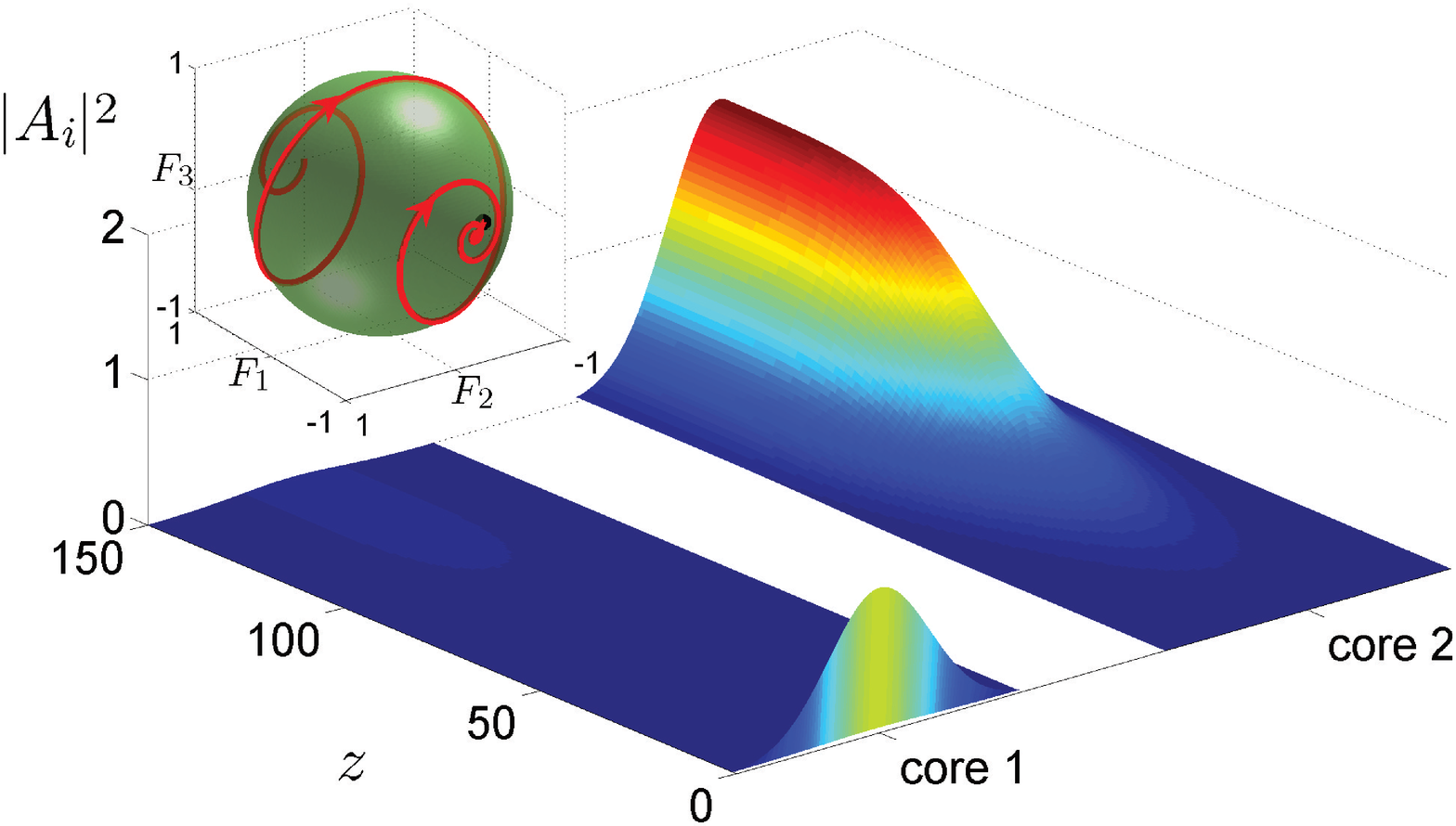}}}
  \subfigure[]{\scalebox{0.2}{\includegraphics{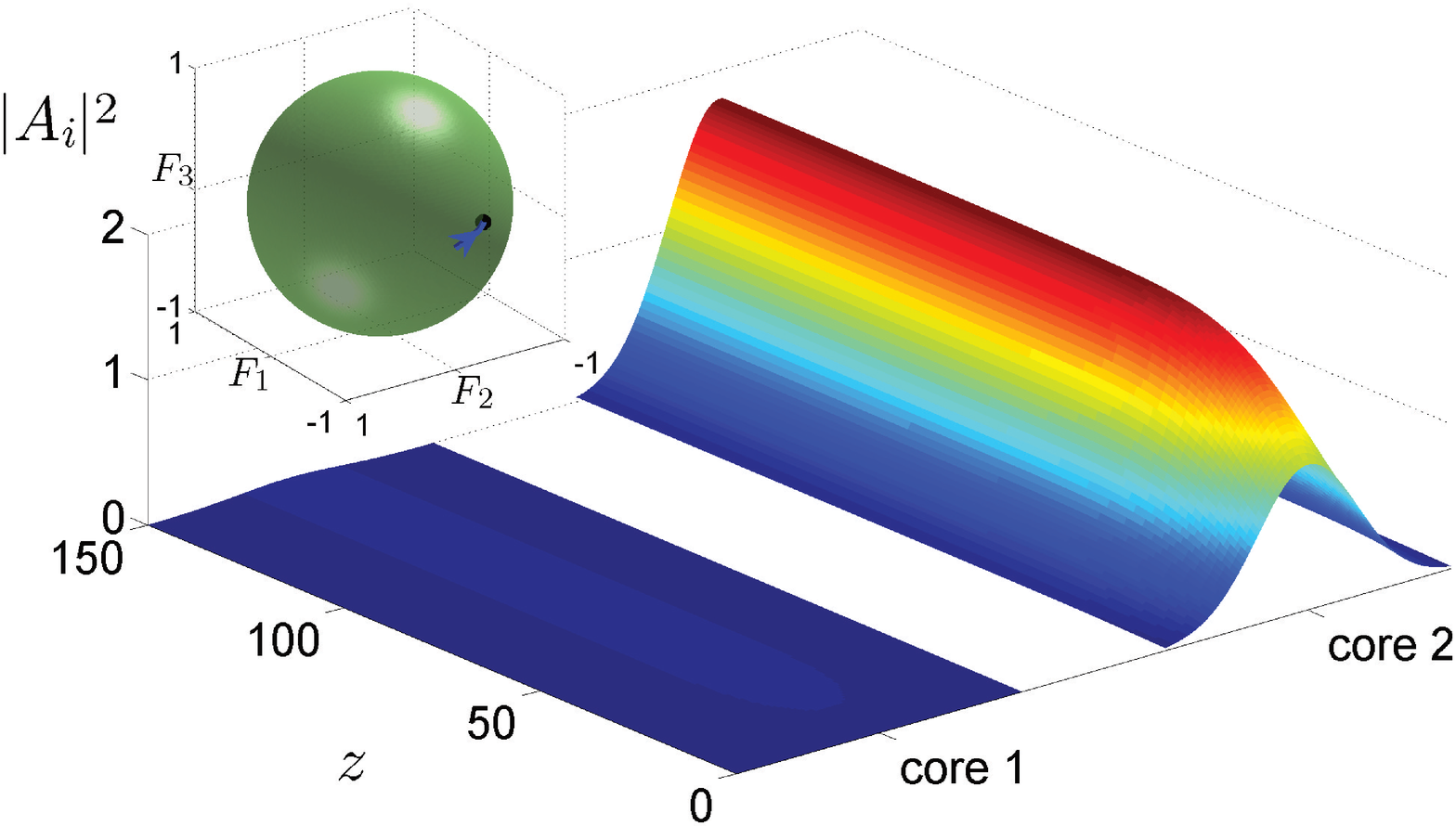}}}
  \caption{Directed power transfer and operation of an Asymmetric Active Coupler as an Optical Isolator. The parameter values are  $\alpha=2$, $X=1$, $K=0.8$, $\Delta=0.95$ and $B=4$. Power is initially launched in the first (a) and second (b) waveguide. The insets depict the evolution of the Stokes vector on a Bloch sphere of unit radius.}
  \end{center}
\end{figure}


\begin{thebibliography}{99}
\bibitem{Lifante} G. Lifante, \textit{Integrated Photonics: Fundamentals} (Wiley, New York,2003).

\bibitem{Jensen_82} S.M. Jensen, IEEE J. Quant. Electron. \textbf{18}, 1580-1583 (1982).

\bibitem{Daino_85} B. Daino, G. Gregori, and S. Wabnitz, J. Appl. Phys. \textbf{58}, 4512-4514 (1985).

\bibitem{Chen_92} Y. Chen, A.W. Snyder, and D.N. Payne, IEEE J. Quant. Electron. \textbf{28}, 239-245 (1992).

\bibitem{Thirstrup_95} C. Thirstrup, IEEE J. Quant. Electron. \textbf{31}, 2101-2106 (1995).

\bibitem{Ramezani_10} H. Ramezani, T. Kottos, R. El-Ganainy, and D.N. Christodoulides, Phys. Rev. A \textbf{82}, 043803 (2010).

\bibitem{Sukhorukov_10} A.A. Sukhorukov, Z. Xu, and Y.S. Kivshar, Phys. Rev. A \textbf{82}, 043818 (2010).

\bibitem{PT_theory} 
C.M. Bender and S. Boettcher, Phys. Rev. Lett. \textbf{80}, 5243-5246 (1998);
R. El-Ganainy, K.G. Makris, D.N. Christodoulides, and Z.H. Musslimani, Opt. Lett. \textbf{32}, 2632-2634 (2007);
K.G. Makris, R. El-Ganainy, D.N. Christodoulides, and Z.H. Musslimani, Phys. Rev. A \textbf{81}, 063807 (2010);
K.G. Makris, Z.H. Musslimani, D.N. Christodoulides, and S. Rotter, Nat. Commun. \textbf{6}, 7257 (2014);
V.V. Konotop, J. Yang, and D.A. Zezyulin, Rev. Mod. Phys. (to appear);

\bibitem{Guo_09} A. Guo, G.J. Salamo, D. Duchesne, R. Morandotti, M. Volatier-Ravat, V. Aimez, G.A. Siviloglou, and D.N. Christodoulides, Phys. Rev. Lett. \textbf{103}, 093902 (2009).

\bibitem{Ruter_10} C.E. Ruter, K.G. Makris, R. El-Ganainy, D.N. Christodoulides, M. Segev, and D. Kip, Nat. Phys. \textbf{6}, 192-195 (2010).

\bibitem{Feng_11} L. Feng, M. Ayache, J. Huang, Y.-L. Xu, M.-H. Lu, Y.-F. Chen, Y. Fainman, and A. Scherer, Science \textbf{333}, 729-733 (2011);
L.Feng,	Y.-L. Xu, W.S. Fegadolli, M.-H. Lu, J.E.B. Oliveira, V.R. Almeida, Y.-F. Chen, and A. Scherer, Nat. Mater. \textbf{12}, 108–113 (2013).

\bibitem{Peng_14} B. Peng, S.K. Ozdemir, F. Lei, F. Monifi, M. Gianfreda, G.L. Long, S. Fan, F. Nori, C.M. Bender, and L. Yang, Nat. Phys. DOI: 10.1038/nphys2927 (2014). 

\bibitem{Bender_13}  N. Bender, S. Factor, J.D. Bodyfelt, H. Ramezani, D.N. Christodoulides, F.M. Ellis, and T. Kottos, Phys. Rev. Lett. \textbf{110}, 234101 (2013).

\bibitem{Barashenkov_14} N.V. Alexeeva, I.V. Barashenkov, K. Rayanov, and S. Flach, Phys. Rev. A 89, 013848 (2014).

\bibitem{Lin_11} Z. Lin, H. Ramezani, T. Eichelkraut, T. Kottos, H. Cao, and D.N. Christodoulides, Phys. Rev. Lett. \textbf{106}, 213901 (2011).

\bibitem{lasers} H. Hodaei, M.-A. Miri, M. Heinrich, D.N. Christodoulides, and M. Khajavikhan, Science, doi:10.1126/science.1258480 (2014);
L. Feng, Z.J. Wong,R.-M. Ma, Y. Wang, and X. Zhang, Science, doi: 10.1126/science.1258479 (2014);
L. Ge and R. El-Ganainy, arXiv:1602.07293 (2016).

\bibitem{Malomed} B.A. Malomed and H.G. Winful, Phys. Rev. E \textbf{53}, 5365-5368 (1996);
J. Atai and B.A. Malomed, Phys. Rev. E \textbf{54}, 4371-4374 (1996);
R. Driben and B.A. Malomed, Opt. Lett. \textbf{36}, 4323-4325 (2011);
Yu. Bludov, V.V. Konotop, and B.A. Malomed, Phys. Rev. A \textbf{87}, 013816 (2013).

\bibitem{Kominis_homogeneous} Y. Kominis, S. Droulias, P. Papagiannis, and K. Hizanidis, Phys. Rev. A \textbf{85}, 063801 (2012);
Y. Kominis, P. Papagiannis, and S. Droulias, Opt. Express \textbf{20}, 18165 (2012).

\bibitem{Musslimani_08} Z.H. Musslimani, K.G. Makris, R. El-Ganainy, and D.N. Christodoulides, Phys. Rev. Lett. \textbf{100}, 030402 (2008);

\bibitem{Kominis_inhomogeneous} Y. Kominis, Opt. Comm. \textbf{334}, 265�272 (2015);
Y. Kominis, Phys. Rev. A \textbf{92}, 063849 (2015).

\bibitem{Aleiner_12}
I.L. Aleiner, B.L. Altshuler, and Y.G. Rubo, Phys. Rev. B \textbf{85}, 121301 (2012);

\bibitem{Rayanov_15}
K. Rayanov, B.L. Altshuler, Y.G. Rubo, and S. Flach, Phys. Rev. Lett. \textbf{114}, 193901 (2015).

\bibitem{Rahmani_16} A. Rahmani and F.P. Laussy, arXiv:1603.05971 (2016).

\bibitem{Barashenkov_13} I.V. Barashenkov, G.S. Jackson, and S. Flach, Phys. Rev. A \textbf{88}, 053817 (2013).

\bibitem{Kevrekidis} P.G. Kevrekidis, D.E. Pelinovsky, and D.Y. Tyugin, J. Phys. A: Math. Theor. \textbf{46} 365201 (2013); 
M. Duanmu, K. Li , R.L. Horne, P.G. Kevrekidis, and N. Whitaker, Phil. Trans. Royal Soc. A, doi:10.1098/rsta.2012.0171 (2013).

\bibitem{Konotop} D.A. Zezyulin and V.V. Konotop, Phys. Rev. Lett. \textbf{108}, 213906 (2012);
D.E. Pelinovsky, D.A. Zezyulin, and V.V. Konotop, J. Phys. A: Math. Theor. \textbf{47}, 085204 (2014).

\bibitem{asymmetric_dimer} A.C. Scott, Phys. Scripta \textbf{42}, 14-18 (1990);
G.P. Tsironis, Phys. Lett. A \textbf{173}, 381-385 (1993). 

\bibitem{PT_integrable} J. Pickton and H. Susanto, Phys. Rev. A \textbf{88}, 063840 (2013);
I.V. Barashenkov, D.E. Pelinovsky, and P. Dubard, J. Phys. A: Math. Theor. \textbf{48}, 325201 (2015).


\end{thebibliography}
\end{document}